\documentclass[conference]{IEEEtran}
\IEEEoverridecommandlockouts
\usepackage{cite}
\usepackage{amsmath,amssymb,amsfonts}
\usepackage{algorithm,algorithmic}
\usepackage{graphicx}
\usepackage{textcomp}
\usepackage{booktabs}
\usepackage{braket}
\usepackage{tabularx}
\def\BibTeX{{\rm B\kern-.05em{\sc i\kern-.025em b}\kern-.08em
    T\kern-.1667em\lower.7ex\hbox{E}\kern-.125emX}}

\usepackage[bookmarks=false, pdfencoding=auto, psdextra, pdfborder={0 0 0}, colorlinks=false, hidelinks]{hyperref}
\usepackage{bm}
\usepackage{mathrsfs}
\usepackage{graphicx}
\usepackage{subfigure}
\usepackage[dvipsnames]{xcolor}
\usepackage{nicefrac}
\usepackage{array}
\usepackage{float}
\usepackage{balance}
\usepackage{multirow} 
\usepackage{quantikz}
\usepackage{tikz}
\usetikzlibrary{arrows.meta,shapes,positioning,fit,backgrounds,calc}
\definecolor{borderpurple}{RGB}{128, 55, 155}
\definecolor{lightpurple}{RGB}{233, 225, 242}
\definecolor{medpurple}{RGB}{201, 176, 242}

\usepackage{adjustbox}
\usepackage[flushleft]{threeparttable} 

\makeatletter
\newcommand*{\transpose}{%
  {\mathpalette\@transpose{}}%
}
\newcommand*{\@transpose}[2]{%
  \raisebox{\depth}{$\m@th#1\intercal$}%
}
\newcommand{\thickhline}{%
    \noalign {\ifnum 0=`}\fi \hrule height 1pt
    \futurelet \reserved@a \@xhline
}
\newcolumntype{"}{@{\hskip\tabcolsep\vrule width 1pt\hskip\tabcolsep}}

\makeatother

\ifCLASSINFOpdf

\else

\fi

\begin{document}
\renewcommand{\ttdefault}{cmtt}
\bstctlcite{IEEEexample:BSTcontrol}

\title{Quantum-Enhanced Reinforcement Learning for Power Grid Security Assessment}

\author{\IEEEauthorblockN{Benjamin M. Peter and Mert Korkali}
\IEEEauthorblockA{\textit{Department of Electrical Engineering and Computer Science}\\ 
\textit{University of Missouri} \\
Columbia, MO 65211 USA \\
e-mail: \{\texttt{bmp792,korkalim\}@missouri.edu}}}

\maketitle

\begin{abstract}
The increasingly challenging task of maintaining power grid security requires innovative solutions. Novel approaches using reinforcement learning (RL) agents have been proposed to help grid operators navigate the massive decision space and nonlinear behavior of these complex networks. However, applying RL to power grid security assessment, specifically for combinatorially troublesome contingency analysis problems, has proven difficult to scale. The integration of quantum computing into these RL frameworks helps scale by improving computational efficiency and boosting agent proficiency by leveraging quantum advantages in action exploration and model-based interdependence. To demonstrate a proof-of-concept use of quantum computing for RL agent training and simulation, we propose a hybrid agent that runs on quantum hardware using IBM’s \textit{Qiskit Runtime}. We also provide detailed insight into the construction of parameterized quantum circuits (PQCs) for generating relevant quantum output. This agent’s proficiency at maintaining grid stability is demonstrated relative to a benchmark model without quantum enhancement using $N-k$ contingency analysis. Additionally, we offer a comparative assessment of the training procedures for RL models integrated with a quantum backend.

\end{abstract}


\section{Introduction}
Current quantum technology is limited to noisy intermediate-scale quantum (NISQ) devices, which are individually insufficient for controlling large-scale power grids. However, by delegating quantum resources for specialized tasks in conjunction with classical grid protection frameworks, quantum computing can be relevant to power system control in the near future.

In the past, classical RL has proven effective in smart grid optimization efforts, as outlined in \cite{vamvakas2023review} for node control, as well as in \cite{ni2019multistage} and \cite{ibrahim2023security} for defending against cyber-physical threats. Reference \cite{alimi2020review} provides an extensive review of machine learning frameworks for grid security applications, highlighting other approaches besides RL.

Previous work has also made initial attempts to leverage quantum solutions for scalable power grid control frameworks. Quantum computing can be applied to alleviate the computational burden of power system operation and planning as discussed in \cite{ganeshamurthy2024next}. This technology is extended explicitly to grid security assessment in \cite{antoli2023quantum}, including an $N-1$ contingency analysis. Tangential problems in power grid research are highlighted in \cite{pareek2024demystifying}, which discusses the use of quantum algorithms for solving power flow problems, and in \cite{ajagekar2019quantum}, which explores the use of quantum computing to optimize the integration of renewables in decarbonizing energy systems. Reference \cite{nikmehr2022quantum} achieves a computational speedup of power grid reliability assessment using a quantum framework rather than traditional Monte-Carlo simulation.

Our motivation for integrating quantum with reinforcement learning (RL) agents specifically pertains to recognizing and mitigating catastrophic grid events. Reference \cite{bienstock2019stochastic} provides a classical approach to grid attack detection using a stochastic defense model. Some of the increasing complexities of modern power system risk assessment are remedied using quantum computing in \cite{hosseini2024modern}. We emphasize assessing the grid’s security in cyberattack scenarios, similar to the contingency analysis performed in \cite{zonouz2013socca} and \cite{zhao2022risk}, in which critical components (i.e., potential targets for malicious attacks) are identified. These classical and quantum approaches to mitigating the impact of extreme grid events prompt our contributions to maintaining power grid stability. Our main contributions are summarized as follows:
\begin{itemize}
    \item A comparative analysis of several quantum-enhanced RL procedures for agent training;
    \item Framework specifications for a hybrid quantum model tuned for power grid security assessment; and
    \item $N-k$ contingency screening analysis of the hybrid quantum model.
\end{itemize}

The remainder of this paper is organized as follows: Section \ref{sec:qrl} provides an overview of quantum RL applied to power grids. Section \ref{sec:qtp} provides a comparative analysis of different training procedures for efficient interaction with the quantum backend. Section \ref{sec:framework} highlights our novel quantum-enhanced hybrid agent model, with numerical results in Section \ref{sec:results} demonstrating its performance. Finally, Section \ref{sec:conclusions} offers concluding remarks and suggests future work.

\section{Quantum Reinforcement Learning}\label{sec:qrl}
We now outline how quantum computing can be used for analyzing power grid security through the interaction between a quantum backend and an agent's simulation environment. We preface this description with some foundational remarks on quantum computing.

\subsection{Background}
Rather than storing information in bits, as in classical computing, quantum computing uses qubits to simultaneously encode classical values, thereby utilizing the principle of superposition. The quantum state of qubits may be altered using various logical gates before being collapsed into a classical outcome upon measurement.

Reference \cite{eskandarpour2020quantum} further outlines the foundational behavior of qubits, including the geometric representation of their state space in a Bloch sphere. This work then proposes a quantum-computing-based algorithm for $N-k$ grid contingency analysis, focusing on efficiently and quickly solving the governing power-flow equations for higher-order ($k>2$) scenarios. We uniquely apply this concept of quantum computing to the training process of grid-based RL agents, while also achieving heightened computational efficiency.

\subsection{Simulation Environment}
RTE France's \textit{Grid2Op} framework \cite{grid2op} is used for training and simulating RL agents for grid operation. This framework uses a Markov decision process (MDP) to act in a given grid environment, specifically a modified version of the IEEE 14-bus test grid. As such, this work represents a unique application of a quantum extension to classical MDPs as proposed in \cite{barry2014quantum}. At its core, \textit{Grid2Op} facilitates the simulation of a given RL agent in an evolving environment with stochastic line loading, achieved through the iterative modification of the action and observation spaces.

\subsection{Quantum Backend}
The simulation interacts with IBM quantum computers using \textit{Qiskit Runtime} \cite{qiskit2024}, which facilitates communication between the \textit{Qiskit} software used in the agents' training procedures and quantum hardware. The training software optimizes efficiency by querying all available IBM quantum computers and using the least busy backend that is operational. \textit{SamplerV2}, a primitive of \textit{Qiskit Runtime}, ran the specified parameterized quantum circuit (PQC) $1{,}024$ times each time it was accessed to generate an averaged output from the quantum hardware. The following subsection and Section \ref{sec:framework} provide details on the operations performed using this hardware, namely the construction of PQCs.

\subsection{Parameterized Quantum Circuits}
\begin{figure}[htbp]
    \centering
    \begin{adjustbox}{max width=\columnwidth}
    \begin{quantikz}[row sep={0.8cm,between origins}, column sep=0.7cm]
        \lstick{$q_0$} & \gate{R_Y(\theta_0)} & \gate{R_Z(\phi_0)} & \qw & \meter{} & \qw \\
        \lstick{$q_1$} & \gate{R_Y(\theta_1)} & \gate{R_Z(\phi_1)} & \qw & \meter{} & \qw \\
        \lstick{$q_2$} & \gate{R_Y(\theta_2)} & \gate{R_Z(\phi_2)} & \qw & \meter{} & \qw \\
        \lstick{$q_3$} & \gate{R_Y(\theta_3)} & \gate{R_Z(\phi_3)} & \qw & \meter{} & \qw \\
    \end{quantikz}
    \end{adjustbox}
    \vspace{-.15in}
    \caption{The four-qubit parameterized quantum circuit with only rotational gates included.}
    \label{fig:rotational-circuit-symbolic}
\end{figure}
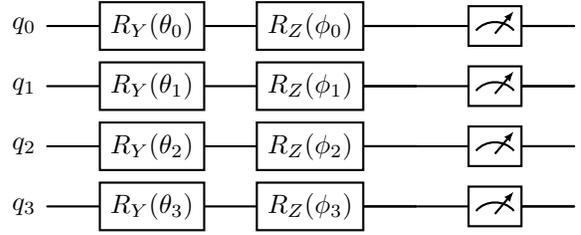

PQCs will be described more rigorously in Section \ref{sec:framework}, but we offer here a brief description of the fundamentals of their use. Reference \cite{Benedetti_2019} outlines the benefits that PQCs offer for machine learning in various applications, namely the increased expressivity they provide for models. Figure \ref{fig:rotational-circuit-symbolic} shows a $4$-qubit circuit which uses rotational gates in an $R_Y \rightarrow R_Z$ ordering to encode a change of basis to the superimposed qubits and facilitate full traversal of the Bloch sphere as described in (\ref{eq:rotation_1}).

\begin{figure}[htbp]
    \centering
    \begin{adjustbox}{max width=\columnwidth}
    \begin{quantikz}
        \lstick{$q_0$} & \gate{H} & \gate{R_Y(\theta_0)} & \gate{R_Z(\phi_0)} & \ctrl{1} & \qw      & \qw      & \qw      & \meter{} & \qw \\
        \lstick{$q_1$} & \gate{H} & \gate{R_Y(\theta_1)} & \gate{R_Z(\phi_1)} & \targ{} & \ctrl{1} & \qw      & \qw      & \meter{} & \qw \\
        \lstick{$q_2$} & \gate{H} & \gate{R_Y(\theta_2)} & \gate{R_Z(\phi_2)} & \qw     & \targ{}  & \ctrl{1} & \qw      & \meter{} & \qw \\
        \lstick{$q_3$} & \gate{H} & \gate{R_Y(\theta_3)} & \gate{R_Z(\phi_3)} & \qw     & \qw      & \targ{}  & \qw      & \meter{} & \qw \\
    \end{quantikz}
    \end{adjustbox}
    \vspace{-.2in}
    \caption{The four-qubit parameterized quantum circuit used for testing the hybrid quantum model.}
    \label{fig:hybrid-circuit-symbolic}
\end{figure}
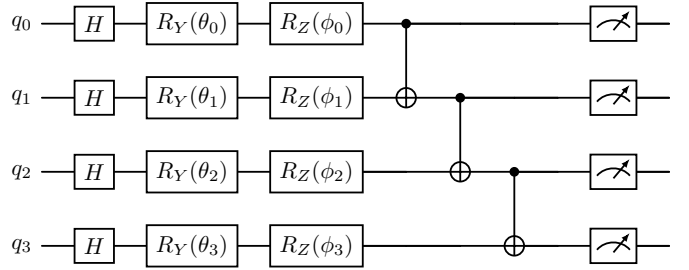

Some initial testing was performed with this relatively basic circuit. However, since it does not utilize the principles of entanglement and superposition, this circuit falls short of the most notable advantages of quantum computing. To remedy this shortcoming, Hadamard and controlled-NOT (CNOT) gates were added to the circuit as seen in Fig. \ref{fig:hybrid-circuit-symbolic}. Section \ref{sec:framework} again describes the use of these gates for the context of this work in more detail, and their definitions are discussed extensively in \cite{crooks2020gates}.

\section{Quantum Training Procedures}\label{sec:qtp}
In this section, we describe the training procedures that govern the interactions between the quantum backend and the simulation environment described in Section \ref{sec:qrl}. We emphasize comparing the computational efficiency and other advantages of three quantum-enhanced procedures: iterative training, cached training, and hybrid training.

\subsection{Iterative Training}
The iterative training procedure executes the PQC at every training step, thus fully leveraging the advantages of heightened expressibility and interdependence modeling. However, reconstructing and executing policy updates using the PQC and transitioning from the quantum backend to the simulation environment at every training step is relatively inefficient. Latency is introduced due to the hardware being accessed and vacated iteratively. For large data sets, these frequent calls using \textit{Qiskit Runtime} can be computationally expensive. These considerations motivate a caching process so that the backend is not accessed for every policy update.

\subsection{Cached Training}
To maximize the efficiency of the quantum training process, a cached training procedure may be used. In this procedure, the PQC is called just once before the quantum output is cached for use throughout the classical portion of the training. In this case, the quantum output,
\begin{equation}
\label{eq:cached}
\mathbf{x}_{\text{quantum}}=[(1-v)\cdot \phi(x)+v\cdot r]\cdot \mathbf{1}_{h},
\end{equation}
is defined from the result of evaluating the PQC, denoted $\phi(x)$, with the cached scalar $r$ being stored after $\phi(x)$ is first evaluated. Here, $v \in \{0,1\}$ is a validity flag for caching and $h$ is the feature dimension. This mitigates the issues of latency and computational expense in the iterative procedure, but fails to provide a significant advantage over traditional RL methods due to the relatively minor contribution from the PQC.

\subsection{Hybrid Training}
\begin{figure*}[htbp]
\centerline{\includegraphics[width=0.95\linewidth]{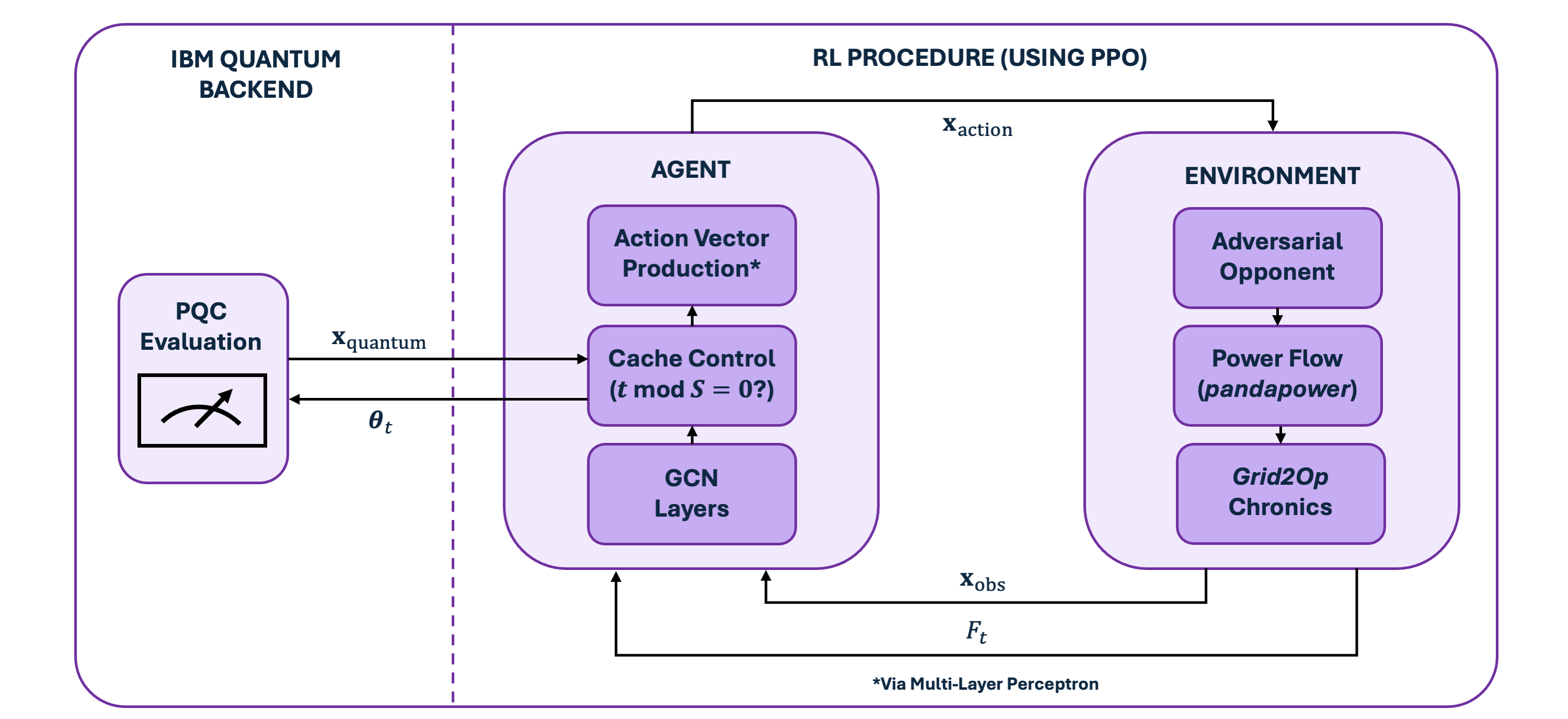}}
\vspace{-.1in}
\caption{Flowchart showing the hybrid training procedure and its interaction with the quantum backend.}
\label{fig:procedures}
\end{figure*}

To balance the efficiency offered by the cached model with the representational power of the iterative model, we finally propose a hybrid training procedure. This procedure caches the quantum circuit output for speed and flexibility, but refreshes after every $S\in\mathbb{N}$ training steps. This combination of caching and iterative procedures is described visually in Fig.~\ref{fig:procedures}.

The hybrid training is the basis for the framework described in Section \ref{sec:framework}, with the hybrid quantum output, $\mathbf{x}_{\text{quantum}}$, defined in detail in (\ref{eq:p_z}).

\section{Proposed Framework}\label{sec:framework}
Initial testing was performed on all three of the previously listed training procedures. However, in this section and Section \ref{sec:results}, we will specifically outline the proposed training framework and simulation results for the hybrid training procedure only. The hybrid model was chosen since it balances training efficiency and effective grid management across the tested scenarios. Note that the framework described here interacts with the \textit{pandapower} \cite{pandapower2018} backend for the \textit{Grid2Op} framework to solve power flow using the Newton-Raphson method, as well as the \textit{IBM Quantum Backend} as described in Section \ref{sec:qrl}. Specific model dimensions are provided in the hyperparameter tables, but the framework discussed in this section is generally applicable to any grid environment.

\subsection{Feature Extraction}\label{sec:gcn}
To train the hybrid model using RL, the agent must extract relevant features from the environment to act upon. The extraction process uses graph convolutional neural networks (GCNs) to dynamically resize the observation space for interaction with PQCs. We model the grid as a graph, $\mathcal{G}=(\mathcal{V},\mathcal{E})$, where $\mathcal{V}$ represents the nodes (buses) and $\mathcal{E}\subseteq \mathcal{V} \times \mathcal{V}$ is the set of edges (transmission lines). Using the GCN convolution function described in \cite{kipf2017semisupervisedclassificationgraphconvolutional} as $\text{Conv}_i$ for layer index $i$, feature extraction is defined as
\begin{subequations}
\begin{align}
\mathbf{L}_1&=\text{ReLU}(\text{Conv}_1(\mathbf{x}_{\text{obs}},\mathcal{E}))\in\mathbb{R}^{|\mathcal{V}|\times h_1}, \label{eq:L_1}
\\
\mathbf{L}_2&=\text{ReLU}(\text{Conv}_2(\mathbf{L}_1,\mathcal{E}))\in\mathbb{R}^{|\mathcal{V}|\times h_2}, \label{eq:L_2}
\end{align}
\end{subequations}
where $\mathbf{x}_{\text{obs}}$ is the observation vector with dimension $d$; and $h_1$ and $h_2$ are the dimensions of hidden layers 1 and 2, respectively. To facilitate generalization of the model across different grid environments, mean pooling is applied to produce output vector $\mathbf{x}_{\text{GCN}}$. The hyperparameters used for the GCN feature extraction process are given in Table \ref{tab:gcn_hyper}.

\begin{table}[h!]
    \centering
    \caption{GCN Hyperparameters}
    \label{tab:gcn_hyper}
    \begin{tabular}{l llll}
        \toprule
        \textbf{Specification}& \textbf{Parameter}& \textbf{Input}& \textbf{Output}&\textbf{Activation}\\
        \midrule
        \multirow{3}{*}{Model Architecture}& Hidden Layer 1&  $64$& $64$&ReLU\\
                                   & Hidden Layer 2&  $64$& $128$&ReLU\\
 & Mean Pooling& $128$& $128$&$-$\\
        \midrule
                                   Quantum Layer& PQC Output&  $-$& $128$&$-$\\
        \bottomrule
    \end{tabular}
    \label{tab:dp_hyperparams}
\end{table}

\subsection{Quantum Enhancement}\label{sec:dual_policy}
We now outline the use of PQCs to produce an output vector, $\mathbf{x}_{\text{quantum}}$, i.e., the output of the PQC given in Fig.~\ref{fig:hybrid-circuit-symbolic}. Here, we describe the functionality of the quantum gates involved. First, each of the $n$ qubits used ($n=4$ for this proof-of-concept case) is initialized in superposition via a Hadamard operation
\begin{equation}
\label{eq:hadamard}
\ket{\psi_H}=\frac{1}{\sqrt{2^n}}\sum_{i\in(0,1)^n}\ket{i}=\ket{+}^{\otimes n},
\end{equation}
where $\psi$ is the quantum state of the circuit's output. This facilitates quantum interference throughout the following operations, a key advantage of this quantum enhancement. Since this is a tensor product of $\ket{+}$, we proceed to apply rotational gates over all $n$ of those qubit states, defined using the unitary operator $U$ on input parameters $\boldsymbol{\theta}\in\mathbb{R}^{2n}$ from the GCN (indexed by $j$) 
\begin{subequations}
\begin{align}
U_j(\boldsymbol{\theta})&=R_Z(\boldsymbol{\theta}_{j+n})\cdot R_Y(\boldsymbol{\theta}_j)\cdot H, \label{eq:rotation_1}\\
U_{\text{encoding}}(\boldsymbol{\theta})&=\bigotimes_{j=1}^{n}U_j(\boldsymbol{\theta}), \label{eq:rotation_2}\\
\ket{\psi_{\text{encoding}}}&=U_{\text{encoding}}(\boldsymbol{\theta})\ket{0}^{\otimes n},
\label{eq:rotation_3}
\end{align}
\end{subequations}
in which $R_Y$ and $R_Z$ rotate the input around the Bloch sphere's $y$ and $z$ axes, respectively. Before measurement, the qubits are entangled using CNOT gates, which introduce interdependency among the parameters. This CNOT chain generates an entanglement operator defined as
\begin{equation}
\label{eq:rotation_2}
U_{\text{entangle}}=\prod_{k=1}^{n-2}\text{CNOT}(q_k,q_{k+1}),
\end{equation}
for qubits denoted by $q$. This results in the final quantum state,
\begin{equation}
\label{eq:rotation_3}
\ket{\psi_{\boldsymbol{\theta}}}=U_{\text{entangle}}\cdot U_{\text{encoding}}(\boldsymbol{\theta})\ket{0}^{\otimes n},
\end{equation}
which is embedded in the expectation value of observable $\mathsf{Z}^{\otimes n}$ as $\bra{\psi_{\boldsymbol{\theta}}}\mathsf{Z}^{\otimes n}\ket{\psi_{\boldsymbol{\theta}}}$.
This expectation value, similar to the theoretical expectation described in \cite{Mitarai_2018}, is represented by the estimator, 
\begin{equation}
\label{eq:estimator}
\mathbb{E}_{\boldsymbol{\theta}}[\mathsf{Z}^{\otimes n}]=\sum_z(-1)^{w(z)}\cdot p_z \in[-1,1],
\end{equation}
where $w_z$ is the Hamming weight of the possible quantum measurement $z$; and $p_z$ is a probability distribution for empirically observing $z$ denoted by 
\begin{equation}
\label{eq:p_z}
p_z=\Pr(z)\text{ for }z \in \{0,1\}^n.
\end{equation}
Then, we finally obtain the quantum feature vector,
\begin{equation}
\label{eq:p_z}
\mathbf{x}_{\text{quantum}}=\left[\left(\frac{\mathbb{E_{\boldsymbol{\theta}}}+1}{2}\right)(x-y)+x\right]\cdot \mathbf{1}_{h_2}\in\mathbb{R}^{h_2},
\end{equation}
by broadcasting the rescaled quantum value across the dimension of $h_2$ from the GCN (for a target range of $[x,y]$). Thus, we produce an output vector, $\mathbf{x}_{\text{quantum}}$, which, when concatenated with $\mathbf{x}_{\text{GCN}}$, produces an action vector,
\begin{equation}
\label{eq:action_vect}
\mathbf{x}_{\text{action}}=(\mathbf{x}_{\text{GCN}}\|\mathbf{x}_{\text{quantum}})\in\mathbb{R}^{2h_2},
\end{equation}
for the agent's reinforcement learning. This action vector now concisely defines features derived from both the grid's topology and the extracted values of the PQC to better model the interdependence of grid features.

\subsection{Reward Function}\label{sec:crf}
Since the reward function is arbitrary and does not directly contribute to the PQC's functionality, we will not outline it in detail here. However, the reward function used for the results given in Section \ref{sec:results} is defined by
\begin{equation}
\label{eq:reward}
F_t=F_s+F_o+F_c+F_a,
\end{equation}
where $F_s$ is a survival bonus; $F_o$ is an overload penalty; $F_c$ is a change-of-topology penalty; and $F_a$ is an action penalty at a given time step $t$.

\subsection{Proximal Policy Optimization} \label{sec:ppo}
The model learns using the proximal policy optimization (PPO) algorithm described in \cite{SchulmanWDRK17}. The PPO gradient step loss at training step $t$ is given by
\begin{equation}
\label{eq:ppo}
\mathcal{L}=\mathbb{E}_t[\min(r_t(\boldsymbol{\theta})\widehat{A}_t,\text{clip}(r_t(\boldsymbol{\theta}),1-\epsilon,1+\epsilon)\widehat{A}_t],
\end{equation}
where the action vector, $\mathbf{x}_\text{action}$, is incorporated into probability ratio, $r_t(\boldsymbol{\theta})$, and the reward factors into the advantage function estimate $\widehat{A}_t$.
The hyperparameters listed in Table \ref{tab:sample_separate_headers} outline the training specifications for PPO interacting with the GCN.
\begin{table}[h!]
    \centering
    \caption{PPO Training Specifications}
    \label{tab:sample_separate_headers}
    \begin{tabular}{ll}
        \toprule
         \textbf{Parameter}& \textbf{Value}\\
        \midrule 
         Gamma& $0.99$\\
                                    Lambda& $0.98$\\
  Learning Rate&$10^{-4}$\\
  Entropy Coefficient&$10^{-3}$\\
                                    Value Function Coefficient& $0.4$\\
        \bottomrule
    \end{tabular}
    \label{tab:dp_hyperparams}
\end{table}
These specifications along with (\ref{eq:ppo}) provide stability as the policy improves during training. Note that PPO's incorporation within \textit{Grid2Op} is made possible using OpenAI's \textit{Gymnasium} API \cite{towers2024gymnasiumstandardinterfacereinforcement} and the \textit{Stable-Baselines3} library \cite{stable-baselines3}.

\subsection{Contingency Analysis}\label{sec:cont_screen}
$N-k$ contingency analysis is used to assess the agent's ability to maintain the stability of the grid. This common form of power system security assessment determines the risk of cascading failures and eventual blackout conditions following the initial outage of $k$ transmission lines. Simulating across contingency scenarios provides a variety of insights into the proficiency of the quantum-enhanced RL agent relative to a benchmark. The comparative performance of agents is measured by the steps survived, reward, and cascading failures, which will be aggregated across all sets of $k$ failures from $\binom{N}{k}$ combinations.

\subsection{Opponent}\label{AA}
Agent simulation is supplemented by a malleable opponent actor, allowing it to reflect various scenarios. We have tested against a variety of opponents to more closely model specific events, such as natural disasters or counterattacks, in which the opponent may also use an RL policy to act intelligently.
We highlight here the opponent used to generate the results in Section \ref{sec:results}. The opponent is given an attack budget $B\in \mathbb{N}$ denoting the maximum total number of lines which it can attack and an attack interval $\tau$ such that the opponent may attack up to $m$ lines at time step $t$ when
\begin{equation}
\label{eq:attack}
t \mod \tau = 0  \text{ and } B>0.
\end{equation}
When attacking the environment, the agent prioritizes the targeting of stressed components, i.e., opponents with high line loading ratio, $\rho_i(t)$, by solving
\begin{equation}
\label{eq:objective}
\max \quad \sum_{i\in \mathcal{S}_t}\rho_i(t)\quad \text{s.t.}\quad|\mathcal{S}_t|\leq B,
\end{equation}
where $\mathcal{S}_t\subseteq \mathcal{L}$ is defined as the top $m$ transmission lines $\mathcal{L}$ which may be attacked, ranked by their respective $\rho_i(t)$. The inclusion of this opponent results in additional stress on both the quantum-enhanced agent and the benchmark agent during the simulation process, better demonstrating their effectiveness against severe grid conditions.

\section{Numerical Results}\label{sec:results}
This section outlines the numerical results of the hybrid quantum agent, denoted \texttt{Quantum Agent}, in an $N-k$ contingency screening security assessment. These results are presented in comparison to an RL model that does not incorporate our hybrid quantum framework but follows the same training and contingency screening procedure otherwise. This model is denoted as \texttt{Benchmark} henceforth.
\begin{table}[htbp]
\caption{Average Time Steps Survived (with a $100$-Step Limit) Across All Contingencies}
\centering
\begin{tabularx}{\linewidth}{|c|>{\centering\arraybackslash}X|>{\centering\arraybackslash}X|>{\centering\arraybackslash}X|}
\hline
& \multicolumn{3}{c|}{\textbf{Number of Initial Failures} $\boldsymbol{(k)}$} \\ \hline
\textbf{Case} & $k=2$ & $k=3$ & $k=4$ \\
\hline
\texttt{Quantum Agent} & $98.95$ & $99.04$ & $99.11$ \\ \hline 
\texttt{Benchmark}     & $28.10$ & $20.07$ & $10.55$ \\ \hline
\end{tabularx}
\label{tab:avg_ts_survived}
\end{table}

\begin{figure}[htbp]
\centerline{\includegraphics[width=1.05\linewidth]{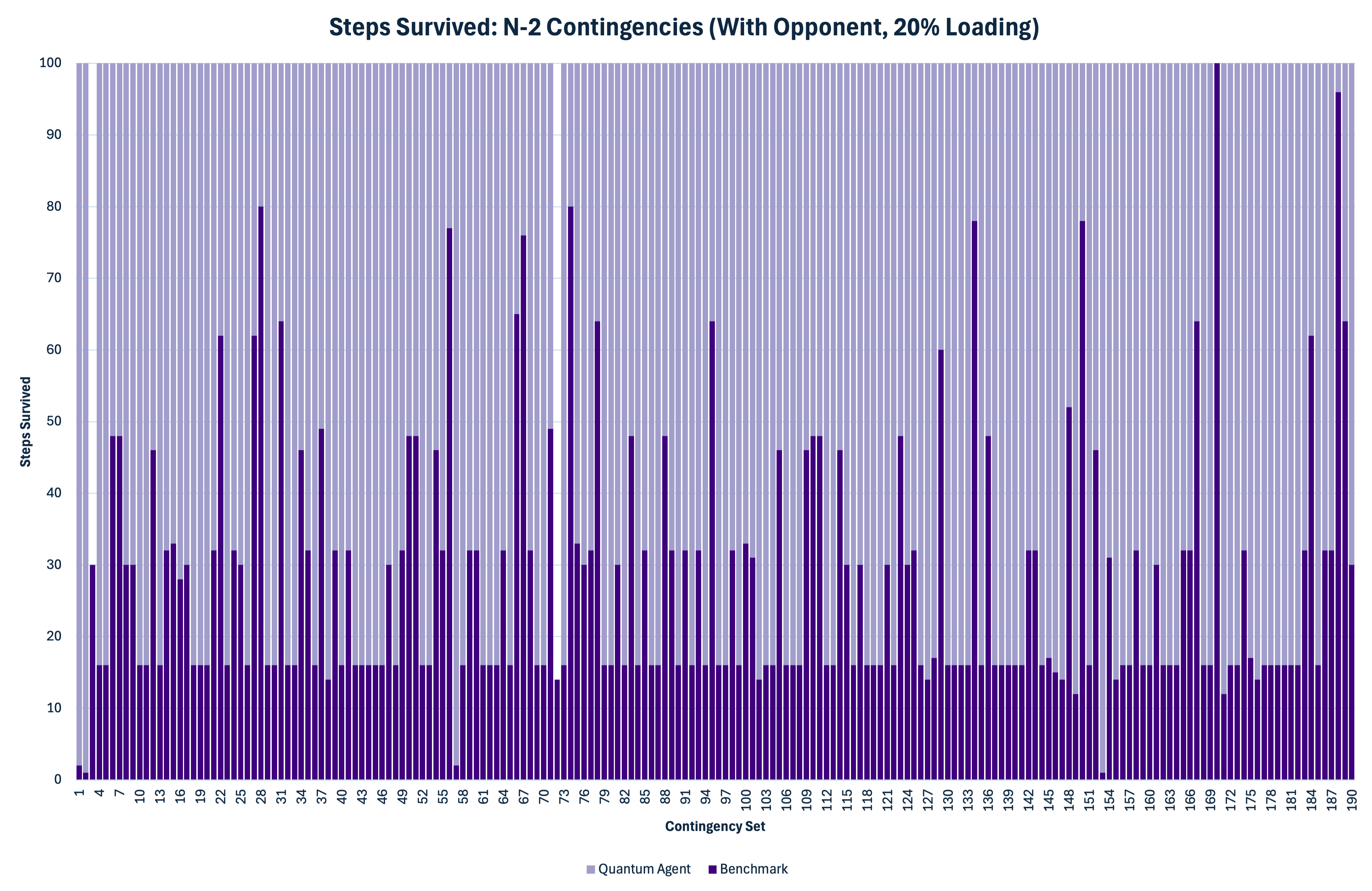}}
\vspace{-.1in}
\caption{Time steps survived for the \texttt{Quantum Agent} and \texttt{Benchmark} cases per $N-2$ contingency set.}
\label{fig:ts_N-2}
\end{figure}

The results given in Table~\ref{tab:avg_ts_survived} show the significant proficiency with which the \texttt{Quantum Agent} acts upon the environment to maximize grid survival time relative to the \texttt{Benchmark}. This comparative performance is visualized in Fig.~\ref{fig:ts_N-2} across $N-2$ contingencies, and highlights the ability of the \texttt{Quantum Agent} to avoid blackout conditions for all $100$ time steps in almost every contingency scenario. The extremely rare cases in which the \texttt{Quantum Agent} could not avoid blackouts were governed by grid-specific initial conditions. This further demonstrates the value of assessing quantum-enhanced RL agents using contingency screening, as critical line combinations may be indicated. Note that these tests included the opponent actor described in Section \ref{sec:framework} along with a $20\%$ increase in line loading to accentuate the agent's relative performance in extreme conditions.

\begin{figure}[htbp]
\centerline{\includegraphics[width=1.05\linewidth]{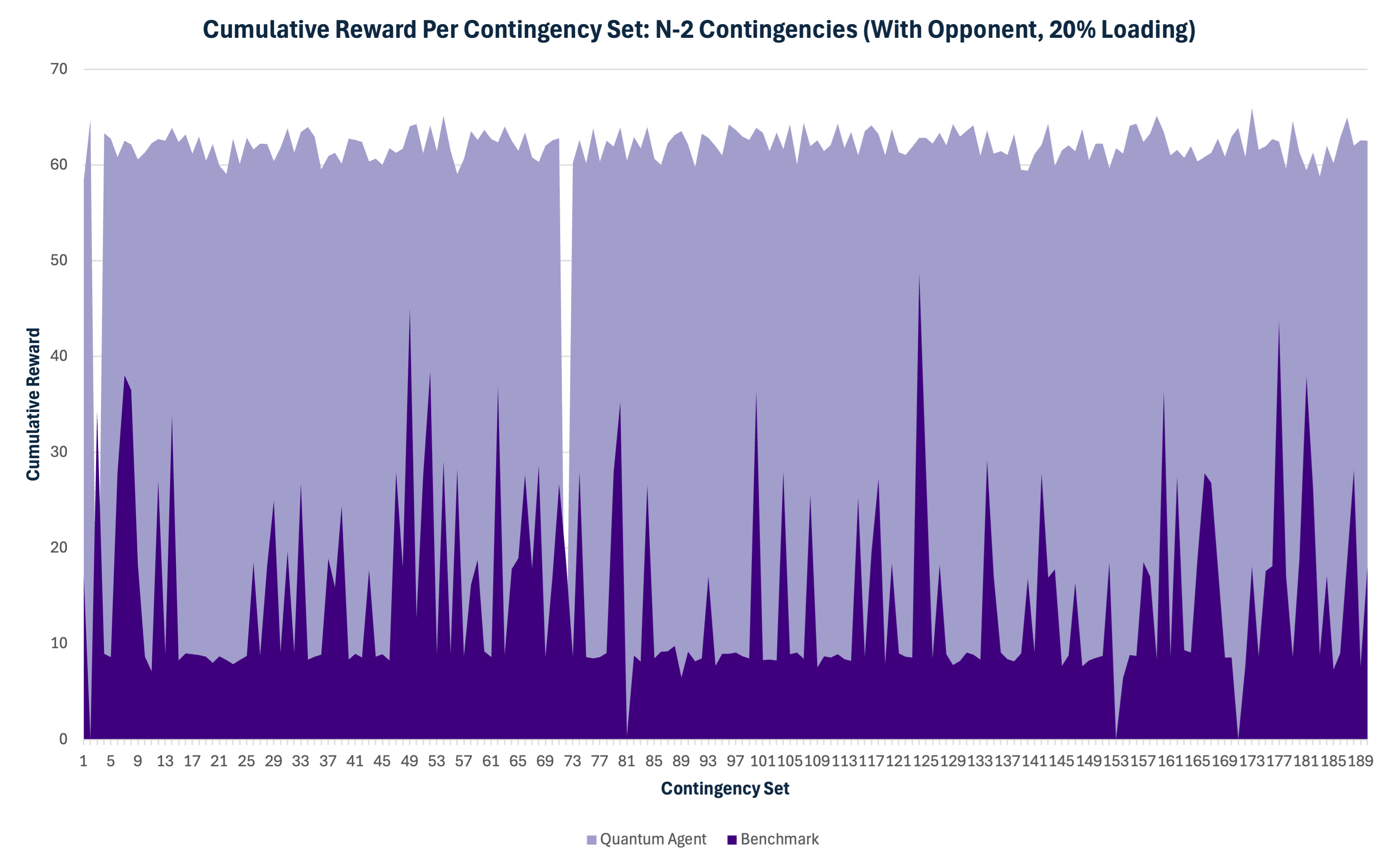}}
\vspace{-.1in}
\caption{Cumulative reward for the \texttt{Quantum Agent} and \texttt{Benchmark} for the $N-2$ contingency set.}
\label{fig:reward}
\end{figure}

Figure \ref{fig:reward} highlights the cumulative reward for the \texttt{Quantum Agent} and \texttt{Benchmark} in each $N-2$ contingency set. This reward encapsulates the proficiency of the \texttt{Quantum Agent} in maintaining stability. It thus signifies its heightened ability to avoid line overloads, frequent topology changes, and other factors that contribute to avoiding blackout conditions. The gap in proficiency at maintaining stability becomes even more pronounced with more extreme conditions, as is evident by the results in Table \ref{tab:avg_ts_survived} for $k=3$ and $k=4$. 

\begin{figure}[htbp]
\centerline{\includegraphics[width=1\linewidth]{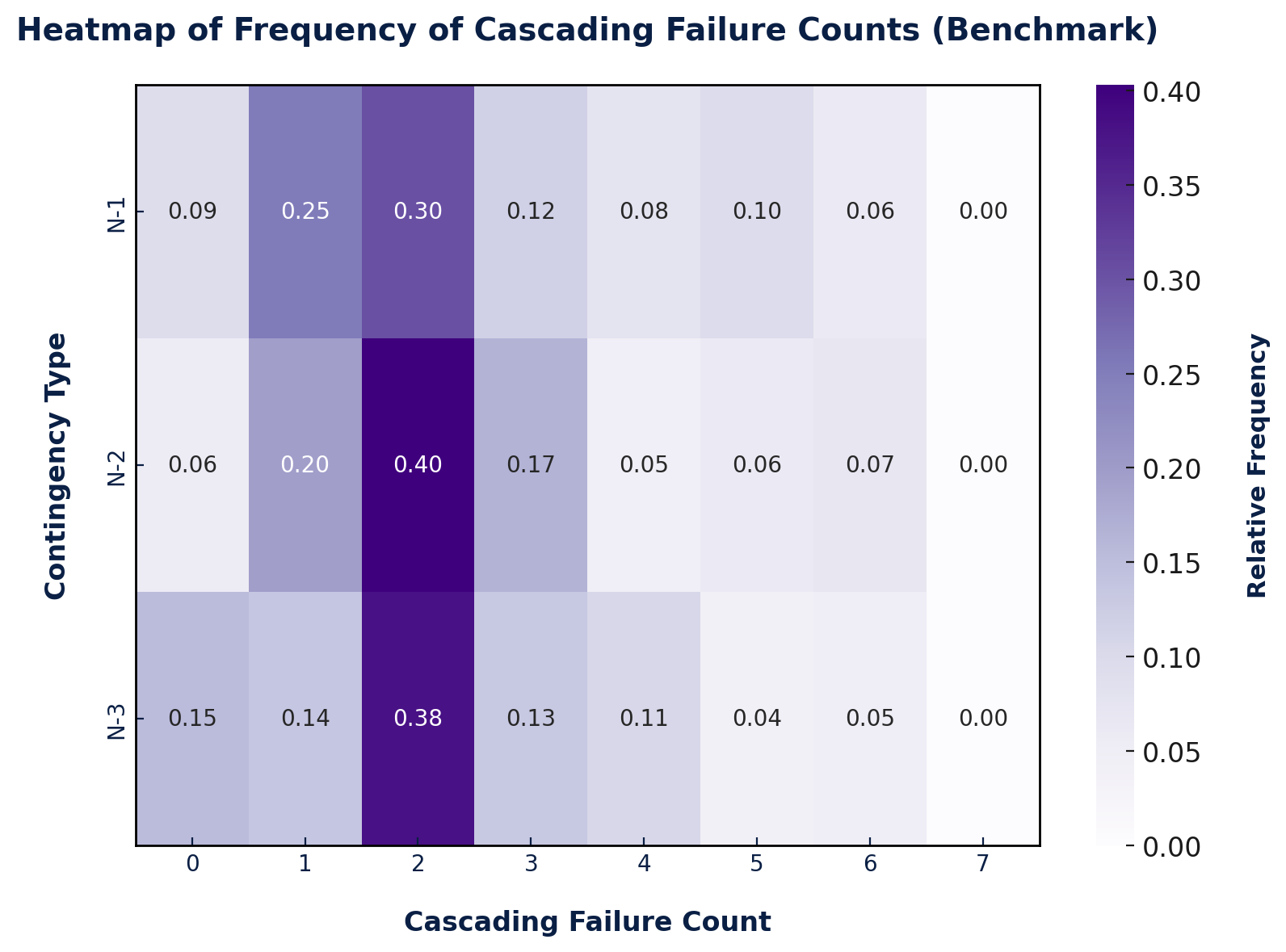}}
\vspace{-.1in}
\caption{Heatmap outlining the relative frequency of cascading failure counts in the \texttt{Benchmark} test for the $N-3$ contingency set.}
\label{fig:heatmap}
\end{figure}

It is interesting to note a key factor in the \texttt{Quantum Agent} outperforming the \texttt{Benchmark} is the avoidance of cascading failures which results from its ability to better model correlated features and explore the action space more effectively. These quantum advantages help it maintain a stable line loading ratio, which is crucial for minimizing the threat of cascades. While these cascades were virtually nonexistent in the \texttt{Quantum Agent} contingency screening, they were quite prevalent in the \texttt{Benchmark} screening. Figure~\ref{fig:heatmap} provides the relative frequency of the number of cascades before blackout conditions were reached.

Note that avoiding cascading failures comes at the risk of a (near-)immediate grid failure. Thus, when any agent is trained with the goal of minimizing the number of cascades, this conflicts with the objective of maintaining survival time and can quickly lead to blackout conditions. This seemed to occur in the rare cases where the \texttt{Quantum Agent} failed to survive all $100$ time steps, and rectifying this issue is an immediate goal for future work.

\section{Conclusion}\label{sec:conclusions}
We have introduced a quantum-enhanced RL agent model for proficient grid stabilization during catastrophic events. Moreover, our hybrid training procedure demonstrates increased efficiency for an RL agent simulation environment interacting with a quantum backend. This work highlights the promise of quantum computing for power systems applications on NISQ devices and fosters optimism for the future of quantum machine learning models for grid control. The agent’s performance relative to the benchmark model highlights the advantages of quantum-enhanced RL models in atypical and extreme conditions (e.g., $k > 2$ with an opponent actor and additional line loading). Thus, this model can act as a risk mitigation strategy amid growing concerns about cyberattacks and natural disasters.

Since the hybrid training procedure facilitated an efficient training process, we aim to scale the contingency analysis simulation to larger power grid test cases. Additionally, we hope to further optimize training adaptability through dynamic PQC construction, which allows for circuit modification during training to adapt to the state of the problem. 

Future work will need to determine how quantum computing can be applied beyond its current highly specialized role in agent training to make large-scale network control feasible through computational speedup. While the practical limitations of NISQ devices make this degree of scaling difficult, we hope that further optimization and tuning of this framework will eventually make extensive quantum integration with the grid a practical reality.

\ifCLASSOPTIONcaptionsoff
\newpage
\fi

\bibliographystyle{IEEEtran}
\bibliography{IEEEabrv,References}

\end{document}